\documentstyle[aps,prl,twocolumn,epsfig]{revtex}
\begin{document}

\title{ Leakage Effect on $J/\psi$ $p_t$ Distributions in
        Different Centrality Bins for Pb - Pb Collisions at E/A = 160 GeV}
\author{Pengfei Zhuang and Xianglei Zhu \\
        Physics Department, Tsinghua University, Beijing 100084, China }
\maketitle

\begin{abstract}
\noindent A transport approach including a leakage effect for
$J/\psi$'s in the transverse phase space is used to calculate the
ratios between the $J/\psi$ transverse momentum distributions in
several centrality bins for Pb-Pb collisions at E/A = 160 GeV.
From the comparison with the CERN-SPS data, where the centrality
is characterized by the transverse energy $E_t$, the leakage
effect is extremely important in the region of high transverse
momentum and high transverse energy, and both the threshold and
the comover models can describe the ratio well for all centrality
bins except the most central one ($E_t < 100$ GeV), for which the
comover model calculation is considerably better than the
threshold one.
\end{abstract}

\noindent ${\bf PACS: 25.75.-q, 12.38.Mh}$\\

Relativistic heavy ion collisions may be the only way to create
the extreme condition for producing a new state of matter ---
Quark-Gluon Plasma (QGP). Among the possible signatures suggested
to identify the existence of QGP, the $J/\psi$ suppression
proposed by Matsui and Satz\cite{matsui} is considered as a direct
one of deconfinement phase transition. The discovery of anomalous
$J/\psi$ suppression\cite{na501} in Pb - Pb collisions in 1996 has
been one of the highlights of the research with relativistic
nuclear collisions at SPS. Since different
models\cite{vogt,gerschel}, with and without the assumption of
QGP, can describe the observed suppression, it is still not yet
clear if the suppression means the discovery of QGP. Recently, the
NA50 collaboration has presented the final analysis of the
$J/\psi$ transverse momentum distribution\cite{na502} in Pb - Pb
collisions at E/A = 160 GeV. The mean squared transverse momentum
$\langle p_t^2\rangle$ as a function of the centrality of the
collisions first increases and then becomes flat when the
centrality increases. The understanding of this saturation of
$\langle p_t^2\rangle$ in central collisions needs the
consideration of time structure\cite{hufner1,hufner2} of the
anomalous suppression. With the idea of leakage, which has already
been considered more than 10 years
ago\cite{matsui,karsch,blaizot1} (c.f. also more recent
works\cite{kharzeev1,shuryak}), the anomalous suppression is not
an instantaneous process, but takes a certain time. During this
time the $J/\psi$'s with high transverse momenta may leak out of
the source of the anomalous suppression. As a consequence, low
$p_t$ $J/\psi$'s are absorbed preferentially, and the mean $p_t$
of the survived $J/\psi$'s in central collisions will not drop
down but become flat. From the comparison of the theoretical
calculations\cite{hufner2} using two different models, the
threshold\cite{blaizot2,blaizot3} and
comover\cite{capella1,capella2} models, with the data, the average
anomalous suppression time $t_A$ is extracted to be $3-4$\ fm/c
for central Pb - Pb collisions. Obviously, the leakage mechanism
affects mainly the high $p_t$ $J/\psi$'s, the mean values $\langle
p_t\rangle < 1.2$\ GeV/c and $\langle p_t^2\rangle < 2$\
(GeV/c)$^2$ for Pb - Pb collisions can not reflect the leakage
effect sufficiently. In order to see the leakage effect more
clearly, we calculate in this paper the $J/\psi$ $p_t$
distributions in different centrality bins, and compare them with
the data of NA50. We will focus our attention on the region
$1<p_t<3$\ GeV/c for central collisions where the leakage effect
and anomalous suppression are both important, and the error bars
of the data are small.

$J/\psi$'s produced in a nucleus-nucleus collision experience
first normal suppression via inelastic $J/\psi - N$ collisions and
show normal broadening of $\langle p_t^2\rangle$ above $\langle
p_t^2\rangle_{NN}$ via gluon rescattering\cite{gerschel}. A gluon
from a projectile nucleon and a gluon from a target nucleon fuse
to form a $J/\psi$ at a point with space coordinates $(\vec s,
z_A)$ in the rest frame of the target A and $(\vec b -\vec s,
z_B)$ in the rest frame of the projectile B, where $\vec b$ is the
impact parameter for the BA collision. On its way out, the
$J/\psi$ experiences the thickness $T_A(\vec s, z_A, \infty)$ and
$T_B(\vec b -\vec s, -\infty, z_B)$ in the nuclei A and B, and is
suppressed with an effective absorption cross section
$\sigma_{abs}$, where the thickness function $T$ is defined as
$T(\vec s,z_1,z_2) = \int_{z_1}^{z_2} dz \rho(\vec s,z)$ with
$\rho_{A,B}$ being the density of nuclei A and B. Before the
fusion, the two gluons traverse thicknesses $T_A(\vec s, -\infty,
z_A)$ and $T_B(\vec b-\vec s, z_B,\infty)$ of nuclear matter in A
and B, and obtain additional transverse momentum via $gN$
collisions. Neglecting effects of formation time\cite{hufner3},
one has after the normal suppression the $J/\psi$ distribution
function $f_N$ in the transverse phase space at given $\vec b$,
\begin{eqnarray}
&& \label{ns} f_N(\vec p_t|\vec b,\vec s) = \int dz_A dz_B
\rho_A(\vec s, z_A)\rho_B
(\vec b-\vec s,z_B)\times \nonumber\\
& &\ \ \ e^{-\sigma_{abs}\left(T_A(\vec s,z_A,\infty)+T_B(\vec
b-\vec s, -\infty, z_B)\right)} {1\over \langle
p_t^2\rangle_N}e^{-p_t^2/\langle p_t^2\rangle_N}
\end{eqnarray}
with
\begin{eqnarray}
&& \label{npt} \langle p_t^2\rangle_N(\vec b,\vec s, z_A,z_B) =
\langle p_t^2\rangle_{NN}+\nonumber\\
&& \ \ \ \ a_{gN}\rho_0^{-1}\left(T_A(\vec s,-\infty,z_A)+T_B(\vec
b-\vec s,z_B,\infty) \right)\ ,
\end{eqnarray}
where we have assumed a Gaussian $p_t$ dependence. The constants
$\sigma_{abs}$ and $a_{gN}$ are usually adjusted to the data from pA
collisions where $J/\psi$'s experience only normal suppression.

Anomalous suppression is attributed to the action on the $J/\psi$
by the mostly baryon free phase of partons and/or hadrons which is
formed after the nuclear overlap. We define $t=0$ as the time when
the normal suppression and normal generation of $\langle
p_t^2\rangle $ have ceased, and describe the time evolution of the
$J/\psi$ density $f(\vec p_t,t|\vec b,\vec s)$ in the transverse
phase space by a transport equation\cite{hufner2}
\begin{equation}
\label{trans}
{\partial f\over \partial t} + \vec v_t\cdot\vec\nabla_s f = -\alpha f
\end{equation}
with the initial condition
\begin{equation}
\label{ini}
f(\vec p_t,0|\vec b, \vec s) = f_N(\vec p_t,\vec b,\vec s)\ .
\end{equation}
The second term on the left handside of the transport equation arises from
the free streaming of the $J/\psi$ with transverse velocity
$\vec v_t = \vec p_t/\sqrt{p_t^2+m_{J/\psi}^2}$, and the loss term is due to
anomalous suppression. The function $\alpha(\vec p_t, t|\vec b,\vec s)$
contains all details about the mechanism of anomalous suppression.

The transport equation (\ref{trans}) can be solved analytically
with the result\cite{hufner2}
\begin{equation}
\label{res} f(\vec p_t, t|\vec b,\vec s) = e^{-\int_0^t
dt'\alpha(\vec p_t,t'|\vec b, \vec s-\vec v_t (t-t'))}f_N(\vec
p_t|\vec b,\vec s-\vec v_t t)\ .
\end{equation}
If the anomalous suppression ceases at time $t_f$, $\alpha(\vec
p_t,t>t_f|\vec b,\vec s) =0$, the observed yield $Y$ as a function
of transverse momentum $p_t$ and transverse energy $E_t$ can be
related to the phase space density $f$ via
\begin{eqnarray}
\label{rel} && Y(p_t,E_t) = \nonumber\\
&&\int d^2\vec b d^2\vec s d^2\vec p_t P(E_t|b)f(\vec p_t, t_f|
\vec b,\vec s)\delta(p_t-\sqrt{p_x^2+p_y^2})\ .
\end{eqnarray}
Here, $P(E_t|b)$ describes the distribution of transverse energy
$E_t$ in events with a given impact parameter $\vec b$. It is
chosen as a Gaussian distribution\cite{blaizot2}
\begin{equation}
\label{gauss}
P(E_t|b) = {1\over \sqrt{2\pi q^2 a N_p(b)}}e^{-{(E_t-\langle E_t\rangle (b))
^2\over 2 q^2 a N_p(b)}}
\end{equation}
with the number of participant nucleons defined as
\begin{eqnarray}
\label{parti} N_p(b) &=& \int d^2\vec s\ n_p(\vec b,\vec s)\ ,\nonumber\\
n_p(\vec b,\vec s) &=& T_A(\vec
s,-\infty,\infty)\left(1-e^{-\sigma_{NN}
T_B(\vec b-\vec s,-\infty,\infty)}\right)\nonumber\\
                   &+& T_B(\vec b -\vec s,-\infty,\infty)\left(1-e^{-
\sigma_{NN}T_A(\vec s,-\infty,\infty)}\right)\ ,\nonumber\\
\langle E_t\rangle (b) &=& qN_p(b)\ ,
\end{eqnarray}
and $a=1.27, q=0.274$\ GeV and $\sigma_{NN} = 32$\
mb\cite{blaizot2} for Pb - Pb collisions at SPS energy.

In order to see clearly the leakage effect which is important for the
$J/\psi$'s with high transverse momenta produced in the events with high
transverse energy, we consider the ratio between the $J/\psi$ $p_t$
distributions in the $E_t$ bin $i$ and in the $E_t$ bin $j$,
\begin{equation}
\label{ratio} R_{i/j} (p_t) = {\int_{\Delta E_t^i} dE_t
Y(p_t,E_t)/N_{\Delta E_t^i}^{DY} \over  \int_{\Delta E_t^j} dE_t
Y(p_t,E_t)/N_{\Delta E_t^j}^{DY}}\ .
\end{equation}
To compare the ratio with the experimental data\cite{na502}, we have
normalized the
distributions to the number of Drell-Yan pairs in the same $E_t$ bins,
\begin{eqnarray}
\label{dy} N_{\Delta E_t^i}^{DY} &=& \int _{\Delta E_t^i} dE_t\int
d^2\vec b d^2\vec s dz_A dz_B P(E_t|b)\times\nonumber\\
&&\rho_A(\vec s,z_A)\rho_B(\vec b-\vec s,z_B)\ .
\end{eqnarray}

Since the physical origin of anomalous suppression is not yet
clear, we calculate the ratio $R$ for two rather different
scenarios: Instantaneous absorption involving a threshold in the
energy density and continuous absorption by comovers.

We begin with the threshold model and use the approach by Blaizot
et. al.\cite{blaizot2} for simplicity. The suppression function in
this approach can be recovered within the transport formalism
(\ref{res}) by setting
\begin{equation}
\label{blaizot} \alpha(\vec p_t, t|\vec b,\vec s) =
\alpha_0\theta(n_p(\vec b,\vec s)-n_c) \delta(t)
\end{equation}
and taking the limit $\alpha_0\rightarrow \infty$. $J/\psi$'s are
totally and instantaneously destroyed if the energy density which
is directly proportional to the participant density $n_p$ is
larger than a critical density, and nothing happens elsewhere.
While the threshold approach successfully describes the data in
the full $E_t$ range of the $J/\psi$ suppression after the only
one free parameter, $n_c$, is adjusted, the predictions for
$\langle p_t^2\rangle(E_t)$ are significantly below the data,
especially in the high $E_t$ region.

It is difficult to understand that the anomalous suppression
happens instantaneously at time $t=0$. The sudden suppression
(\ref{blaizot}) was changed to a continuous suppression between
times $t_0$ and $t_1$\cite{hufner2},
\begin{equation}
\label{change}
\alpha(\vec p_t,t|\vec b,\vec s) = {\alpha_0\over t_1 - t_0}
\theta(n_p(\vec b,\vec s)-n_c)\theta(t_1 - t)\theta(t-t_0)\ .
\end{equation}
From the comparison\cite{hufner2} of the yield $Y(E_t)$ and mean
transverse momentum $\langle p_t^2\rangle(E_t)$ with the data, the
average anomalous suppression time $t_A$ was found to be a
function of $E_t$. It increases from $0$ to about $4$\ fm/c when
$E_t$ increases from $0$ to the maximum ($\sim 140$\ GeV) for Pb -
Pb collisions.

\begin{figure}
\centerline{\includegraphics{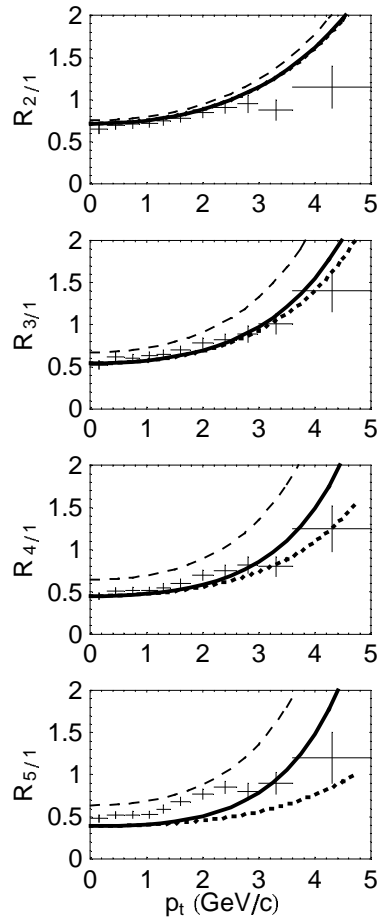}} \caption{ The ratio
between the $J/\psi$ $p_t$ distributions in the $E_t$ bin $i \
(i=2,\ 3,\ 4,\ 5)$ and in the first $E_t$ bin ($i=1$). The
distributions are normalized to the Drell-Yan number in the same
$E_t$ bins. The five $E_t$ bins correspond to the $E_t$ range
$\Delta E_t = [5,34],\ [34,60],\ [60,80],\ [80,100],\ [100,140]$.
The data are from NA50. Dashed, dotted and solid lines are
calculated within the threshold model. The dashed lines show the
result of normal suppression alone, the dotted lines include the
anomalous suppression but without leakage effect, and the solid
lines include also the leakage effect with the average anomalous
suppression time $t_A = 0,\ 1,\ 2,\ 3,\ 4\ fm/c$ corresponding to
the $E_t$ bin $i=1,\ 2,\ 3,\ 4,\ 5$, respectively. } \label{figb}
\end{figure}

With the parameters $\sigma_{abs} = 6.4$\ mb and $n_c = 3.75$\
fm$^{-2}$\cite{blaizot2} for the suppression and $\langle
p_t^2\rangle_{NN} = 1.11$\ (GeV/c)$^2$ and $a_{gN} = 0.081$\
(GeV/c)$^2$ fm$^{-1}$\cite{na502} for the mean transverse
momentum, we calculated the ratios $R_{i/j}(p_t)$. We also
accounted for the transverse energy fluctuations\cite{blaizot2}
which have been shown to be significant for the explanation of the
sharp drop of $J/\psi$ suppression in the domain of very large
$E_t$ values, by replacing $n_p$ by ${E_t\over \langle E_t\rangle}
n_p$ where $\langle E_t\rangle$ is the mean transverse energy at
given $b$. For the comparison with the data\cite{na502} of Pb - Pb
collisions, we consider five $E_t$ bins with $\Delta E_t^i =
[5,34],\ [34,60],\ [60,80],\ [80,100],\ [100,140]$ GeV
corresponding to $i=1,\ 2,\ 3,\ 4,\ 5$, respectively. The
theoretical results $R_{i/1}(p_t)$ with $i=2,\ 3,\ 4,\ 5$ together
with the data are shown in Fig.~\ref{figb}. The dashed lines
indicate the results with normal suppression only. Since the
nuclear absorption deviates from the data and becomes saturated
above $E_t\sim 40$\ GeV\cite{vogt,gerschel}, the calculated ratios
$R_{i/1}$ are always above the data for any values of $p_t$ and in
any $E_t$ bin. The dotted lines are calculated with the original
Blaizot approach (\ref{blaizot}) without considering leakage
effect ($\vec v_t = 0$ in (\ref{res})), and the solid lines are
obtained by taking the average anomalous suppression time $t_A =
0,\ 1,\ 2,\ 3,\ 4$ fm corresponding to the $E_t$ bins $i=1,\ 2,\
3,\ 4,\ 5$, respectively. It is clear that the leakage effect
becomes more and more important when the anomalous suppression
time increases and/or the transverse momentum $p_t$ increases.
While the instantaneous threshold approach, namely the dotted
lines, can describe the data in the region $p_t < 3$\ GeV/c when
$E_t$ is not high enough, the calculated ratio $R_{5/1}$ deviates
from the data clearly. Due to the leakage effect reflected in the
transport approach (\ref{res}), high transverse momentum
$J/\psi$'s may escape the anomalous suppression. This is the
reason why the solid line is always above the dotted line in any
$E_t$ bin. When the degree of centrality is low, the system is
difficult to reach the critical density $n_c$, the anomalous
suppression is not important, and then the leakage effect is weak.
With increasing centrality, more anomalous suppression happens and
more $J/\psi$'s leak out of the suppression region. For the $E_t$
bins $1$ and $2$, almost no leakage happens, the solid and dotted
lines coincide for $R_{2/1}$. For $R_{3/1}$ the leakage is still
not important. However, when the centrality increases further, the
leakage effect leads to a remarkable difference between the solid
and dotted lines. For $R_{4/1}$ and $R_{5/1}$ the leakage improves
considerably the threshold approach. Due to the big error bars of
the data above $p_t$ around $3$\ GeV/c, it is difficult to judge
the leakage effect in the very high $p_t$ region from the
comparison with the data.

We now turn to the comover model. The comoving partons and/or
hadrons lead to a continuous $J/\psi$ absorption of long duration
due to inelastic collisions with the comoving particles. As a
representative model, we use the approaches by Capella et
al.\cite{capella1,capella2} and Kharzeev et.
al.\cite{kharzeev2}.The absorption term $\alpha$ in the transport
equation (\ref{trans}) takes the form\cite{hufner2}
\begin{equation}
\label{comover} \alpha(\vec p_t,t|\vec b,\vec s) =
\sigma_{co}{n_c(\vec b,\vec s)\over t} \theta\left({n_c(\vec
b,\vec s)\over n_f}t_0 - t\right)\theta(t-t_0)\ ,
\end{equation}
where $\sigma_{co}$ is the cross section of $J/\psi$-comover
interaction, $n_c$ the comover density at initial time $t_0$. The
absorption

\begin{figure}
\centerline{\includegraphics{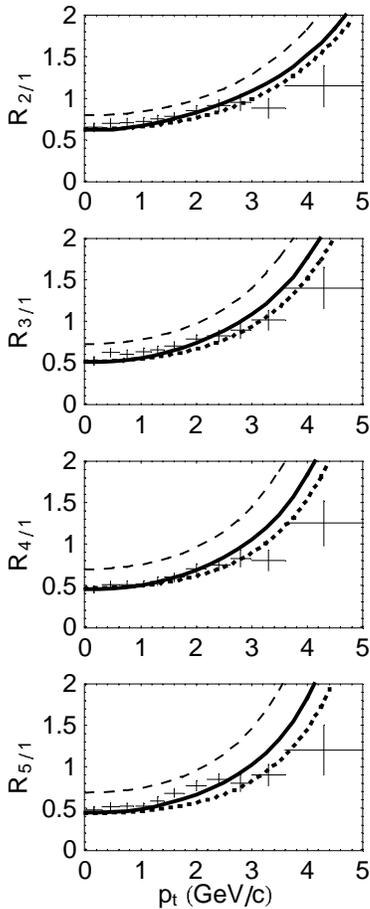}} \caption{ The ratio
between the $J/\psi$ $p_t$ distributions in the $E_t$ bin $i\
(i=2,\ 3,\ 4,\ 5)$ and in the first $E_t$ bin ($i=1$). The
distributions are normalized to the Drell-Yan number in the same
$E_t$ bin. The five $E_t$ bins correspond to the $E_t$ range
$\Delta E_t = [5,34],\ [34,60],\ [60,80],\ [80,100],\ [100,140]$.
The data are from NA50.  Dashed lines (with only normal
suppression), dotted lines (with anomalous suppression but without
leakage effect) and solid lines (with anomalous suppression and
leakage effect) are calculated within the comover model. }
\label{figc}
\end{figure}

\noindent by comovers starts at $t_0$ and ends at $t_1$ when the
comover density $n_c(\vec b,\vec s)t_0/t_1$ has reached a value
$n_f$ independent of $\vec b$ and $\vec s$. Unlike the threshold
scenario for which we have to introduce a time structure
(\ref{change}) to discuss the leakage effect, the comover scenario
contains a definite time structure for anomalous suppression and
it is not necessary to introduce any new parameter. Taking into
account the transverse energy fluctuations\cite{capella1} by
replacing $n_c$ by ${E_t\over \langle E_t \rangle}n_c$, and the
transverse energy loss\cite{capella2} induced by the $J/\psi$
trigger by rescaling $\langle E_t \rangle$ by a factor ${n_p -2
\over n_p}$, which have been shown to be significant for the
explanation of the sharp decrease of $J/\psi$ suppression at $E_t
> 100$ GeV, and using the parameters
$n_{co}(\vec b,\vec s) = 1.5\ n_p(\vec b,\vec s), t_0 = 1$\ fm,
$n_f = 1$\ fm$^{-2}$, $\sigma_{abs} = 4.5$\ mb and $\sigma_{co} =
1$\ mb\cite{capella1,kharzeev2}, the calculated $\langle
p_t^2\rangle(E_t)$\cite{hufner2} in the transport approach
including leakage effect agrees well with the data, and the
average time for comover action is $t_A = 3.5$\ fm/c. With the
same parameters, the calculated ratios $R_{i/1}$ for $i=2,3,4,5$
are shown in Fig.~\ref{figc}. Again the result with only normal
suppression deviates from the data. Compared with the threshold
scenario, the prediction with comover absorption but without
considering leakage effect has a better agreement with the data.
Since the comover absorption is a continuous process, and it can
happen in the events with low centrality, the leakage effect
starts to change the suppression ratio at low $E_t$ and is almost
$E_t$ independent. It is clear that the leakage effect leads to a
better agreement between the comover model and the data. From the
comparison with the threshold scenario, the result of the comover
scenario looks more close to the experimental data.

With the general transport equation proposed in ref.\cite{hufner2}
which describes the time evolution of the $J/\psi$'s during
anomalous suppression including leakage effect, we calculated
within the threshold and comover models the ratio $R_{i/1}(p_t)$
between the $J/\psi$ $p_t$ distribution in the higher $E_t$ bin
$i$ and the same distribution in the lowest $E_t$ bin. Our
motivation is to study the leakage effect in high $p_t$ and high
$E_t$ region where the anomalous suppression is most important. We
found that in this region the leakage effect leads to a better
agreement with the data for the two models. The results support
the idea\cite{hufner2} that the anomalous suppression happens on
average at $3-4$\ fm/c after the normal suppression ends. From the
comparison of the two very different models, our calculation with
the mechanism of continuous comover action over a comparably long
time can better describe the ratios for all $E_t$ bins.

\vspace{0.3in}

\noindent {\bf \underline{Acknowledgments:}} We are grateful to
Prof. Joerg Huefner for many stimulating discussions and careful
reading of the manuscript. This work has been supported by the
grants NSFC19925519, 10135030, and G2000077407.


\begin{thebibliography}{20}

\bibitem{matsui}    T. Matsui, H. Satz,
                    Phys. Lett. {\bf B 178} (1986) 416.
\bibitem{na501}     M.C. Abreu et al., NA50 Collaboration,
                    Nucl. Phys. {\bf A 610} (1996) 404c.
\bibitem{vogt}      R. Vogt,
                    Phys. Rep. {\bf 310} (1999) 197.
\bibitem{gerschel}  C. Gerschel, J. H\"ufner,
                    Ann. Rev. Nucl. Part. Sci. {\bf 49} (1999) 255.
\bibitem{na502}     M.C. Abreu et al., NA50 Collaboration,
                    Phys. Lett. {\bf B 499} (2001) 85.
\bibitem{hufner1}   J. H\"ufner, P. Zhuang,
                    Phys. Lett. {\bf B 515} (2001) 115.
\bibitem{hufner2}   J. H\"ufner, P. Zhuang,
                    nucl-th/0208004, accepted by Phys. Lett. {\bf B}.
\bibitem{karsch}    F. Karsch, R. Petronzio,
                    Z. Phys. {\bf C 37} (1988) 627.
\bibitem{blaizot1}  J.P. Blaizot, J.Y. Ollitrault,
                    Phys. Lett. {\bf B 199} (1987) 627.
\bibitem{kharzeev1} D. Kharzeev, M. Nardi, H. Satz,
                    Phys. Lett. {\bf B 405} (1997) 14.
\bibitem{shuryak}   E. Shuryak, D. Jeaney,
                    Phys. Lett. {\bf B 430} (1998) 37.
\bibitem{blaizot2}  J.P. Blaizot, P.M. Dinh, J.Y. Ollitrault,
                    Phys. Rev. Lett. {\bf 85} (2000) 4010.
\bibitem{blaizot3}  J.P. Blaizot, J.Y. Ollitrault,
                    Phys. Rev. Lett. {\bf 77} (1996) 1703.
\bibitem{capella1}  A. Capella, E.G. Ferreiro and A.B. Kaidalov,
                    Phys. Rev. Lett. {\bf 85} (2000) 2080.
\bibitem{capella2}  A. Capella, A.B. Kaidalov and D. Sousa,
                    Phys. Rev. {\bf C65} (2002)054908.
\bibitem{hufner3}   J. H\"ufner, B.Z. Kopeliovich,
                    Phys. Rev. Lett. {\bf 76} (1996) 192.
\bibitem{kharzeev2} D. Kharzeev, C. Louren{\c c}o, M. Nardi, H. Satz,
                    Z. Phys. {\bf C74} (1997) 307.

\end{thebibliography}
\end{document}